\documentclass[aps,preprint,letterpaper,showpacs,prd]{revtex4}
\usepackage{amsfonts}
\usepackage{amsmath}
\usepackage{amssymb}
\usepackage{graphicx}
\usepackage[pdftex,colorlinks=true,linkcolor=blue,citecolor=blue,filecolor=blue,urlcolor=black
]{hyperref}

\providecommand{\U}[1]{\protect\rule{.1in}{.1in}}

\input{tcilatex}
\begin{document}

\title{$5$-dimensional Einstein-Chern-Simons gravity and a scalar-tensor
theory in $4$-dimensions}
\author{S. Lepe$^{1a}$, C. Riquelme$^{2b}$ \ and P. Salgado$^{3c}$}
\affiliation{$^{1}$Instituto de F\'{\i}sica, Pontificia Universidad Cat\'{o}lica de
Valpara\'{\i}so, Avda. Brasil 2950, Vapara\'{\i}so, Chile\\
$^{2}$Departamento de F\'{\i}sica, Universidad de Concepci\'{o}n, Casilla
160-C, Concepci\'{o}n, Chile\\
$^{3}$Instituto de Ciencias Exactas y Naturales, Facultad de Ciencias,
Universidad Arturo Prat, Avda. Artuto Prat 2120, Iquique, Chile.}

\begin{abstract}
It is shown that the so-called $5$-dimensional generalized Poincar\'{e}%
-Chern-Simons gravity compactified to $4$- dimensions is equivalent to 
\textbf{a} scalar-tensor gravity theory without kinetic term, which can be
cast into a $f\left( R,\phi \right) $ theory type that in a particular case
lead to a $f(R)$ theory. We study black hole solutions of the corresponding $%
4$-dimensional field equations. It is found that for a spherically symmetric
metric such equations lead to two solutions. One such solution is a
non-causal one. The second solution, depending on an arbitrary constant $A$,
is given by a\textbf{\ }spacetime with a cosmological constant inversely
proportional to the square of the compactification radius. If the constant $A
$ is positive, we find a Schwarzschild-de Sitter black hole. If $A$ is
negative, the solution can be understood as a white hole solution which is
obtained applying to the solution with $A>0$ the discrete coordinate
transformation $PT$ accompanied by the transformation $A\rightarrow -\tilde{A%
}$, with $\tilde{A}>0$, corresponding to a transformation known as mass
reversal.

$a$ \ samuel.lepe@pucv.cl $\ $

$b$ \ ceriquelme@udec.cl

$c$ patsalgado@unap.cl
\end{abstract}

\maketitle

\section{\qquad Introduction}

In \cite{salg1} was shown that the standard $5$-dimensional General
Relativity (GR), without cosmological constant, can be obtained from
Chern-Simons gravity theory for a certain Lie algebra $\mathcal{B}$ \cite%
{salg2}, \cite{salg3}. The Chern-Simons Lagrangian for the $\mathcal{B}$
algebra is given by

\begin{equation}
L_{\mathrm{ChS}}^{(5)}=\alpha _{1}l^{2}\varepsilon
_{abcde}R^{ab}R^{cd}e^{e}+\alpha _{3}\varepsilon _{abcde}\left( \frac{2}{3}%
R^{ab}e^{c}e^{d}e^{e}+2l^{2}k^{ab}R^{cd}T^{\text{ }e}+l^{2}R^{ab}R^{cd}h^{e}%
\right) ,  \label{26}
\end{equation}%
where $\alpha _{1}$ and $\alpha _{3}$ are arbitrary constants and $l$ can be
interpreted as a coupling constant that characterizes different regimes
within the theory.

The field content induced by the $\mathcal{B}$ algebra includes the vielbein 
$e^{a}$, the spin connection $\omega ^{ab}$ and two extra bosonic fields $%
h^{a}$ and $k^{ab}$. From (\ref{26}) we can see that it is possible to
recover the $5$-dimensional Einstein gravity theory from a Chern-Simons
gravity theory in the limit where the coupling constant $l$ equals to zero
while keeping the effective Newton's constant fixed.

In the presence of matter described by the langragian $%
L_{M}=L_{M}(e^{a},h^{a},\omega ^{ab})$, in Ref. \cite{salga1} was found that
the field equations admit black hole-type solutions for a spherically
symmetric metric.

On the other hand, in Ref. \cite{bra1}, \cite{bra2} was found that if one
replaces a Randall-Sundrum type metric \cite{randall} \cite{randall1} in the
Einstein-Chern-Simons (EChS) gravity Lagrangian (\ref{26}), the following
action can be obtained

\begin{eqnarray}
\tilde{S}[\tilde{e},\tilde{h}] &=&\int_{\Sigma _{4}}\tilde{\varepsilon}%
_{mnpq}\left( -\frac{1}{2}\tilde{R}^{mn}\tilde{e}^{p}\tilde{e}^{q}\right. + 
\notag \\
&&\left. +K\tilde{R}^{mn}\tilde{e}^{p}\tilde{h}^{q}-\frac{K}{4r_{c}^{2}}%
\tilde{e}^{m}\tilde{e}^{n}\tilde{e}^{p}\tilde{h}^{q}\right) ,  \label{s3}
\end{eqnarray}%
which is a gravity action with a cosmological constant for a $4$-dimensional
brane embedded in the $5$-dimensional spacetime of the so-called EChS
gravity.\textbf{\ }$\tilde{\varepsilon}_{mnpq}$\textbf{, }$\tilde{e}^{m}$, $%
\tilde{R}^{mn}$\textbf{\ }and\textbf{\ }$\tilde{h}^{m}$\textbf{\ }represent,
respectively, the\textbf{\ }$4$\textbf{-}dimensional versions of the
Levi-Civita symbol,\textbf{\ }the vielbein, the curvature form, a\textbf{\ }%
matter field and $r_{c}$\textbf{\ }is the so-called compactification radius. 
$K$ is a constant related to $\alpha _{3}$, $l$ and $r_{c}$. It is of
interest to note that the field $h^{a}$ gives rise to a form of positive
cosmological constant which appears as a consequence of modifications of the
Poincar\'{e} symmetries, carried out through the expansion procedure.\textbf{%
\ }The corresponding version of Eq. (\ref{s3}) in tensor language Ref. \cite%
{bra1}, \cite{bra2} is given by

\begin{equation}
\tilde{S}[\tilde{g},\tilde{h}]=\int d^{4}\tilde{x}\sqrt{-\tilde{g}}\left[ 
\tilde{R}+2K\left( \tilde{R}\tilde{h}-2\tilde{R}_{\text{ }\nu }^{\mu }\tilde{%
h}_{\text{ }\mu }^{\nu }\right) -\frac{3K}{2r_{c}^{2}}\tilde{h}\right] ,
\label{31t'}
\end{equation}%
where $K=4\pi \alpha _{3}l^{2}/r_{c}$, so that when $l\longrightarrow 0$
then $K\longrightarrow 0$ and the actions (\ref{s3}) and (\ref{31t'})
becomes the $4$-dimensional Einstein-Hilbert action. \ 

An interesting observation that if we consider a maximally symmetric
spacetime, then the equation 13.4.6 of Ref. \cite{weinberg} allows us to
write the field\textbf{\ }$\tilde{h}_{\mu \nu }$\textbf{\ }as\textbf{\ }

\begin{equation}
\tilde{h}_{\mu \nu }=\frac{1}{4}\tilde{F}(\tilde{\phi})\tilde{g}_{\mu \nu },
\label{66t}
\end{equation}%
where $\tilde{F}$ is an arbitrary function of an $4$-scalar field $\tilde{%
\phi}=$ $\tilde{\phi}(\tilde{x})$.\ This means

\begin{equation}
\tilde{R}_{\text{ }\nu }^{\mu }\tilde{h}_{\text{ }\mu }^{\nu }=\frac{1}{4}%
\tilde{F}(\tilde{\phi})\tilde{R}\text{ \ \ },\text{\ \ }\tilde{h}=\tilde{h}%
_{\mu \nu }\tilde{g}_{\text{ }}^{\mu \nu }=\tilde{F}(\tilde{\phi}),
\end{equation}%
so that the action (\ref{31t'}) takes the form 
\begin{equation}
\tilde{S}[\tilde{g},\tilde{\phi}]=\int d^{4}\tilde{x}\sqrt{-\tilde{g}}\left[ 
\tilde{R}+K\tilde{R}\tilde{F}(\tilde{\phi})-\frac{3K}{2r_{c}^{2}}\tilde{F}(%
\tilde{\phi})\right] ,  \label{32t}
\end{equation}%
which corresponds to an action for the $4$-dimensional gravity coupled
non-minimally to the scalar field. Note that this action has the form $%
\tilde{S}=\tilde{S}_{g}+\tilde{S}_{g\phi }+\tilde{S}_{\phi },$ where, $%
\tilde{S}_{g}$ is a pure gravitational action term, $\tilde{S}_{g\phi }$ is
a non-minimal interaction term between gravity and the scalar field, and $%
\tilde{S}_{\phi }$ represents the action for a kind of scalar field
potential. In order to write down the action in the usual way, we define the
constant $\varepsilon $ and the potential $V(\tilde{\phi})$ as (removing the
symbols $\sim $ in (\ref{32t})) 
\begin{equation}
\varepsilon =\frac{4\kappa r_{c}^{2}}{3}\text{ \ \ },\text{\ \ }V(\phi )=%
\frac{3K}{4\kappa r_{c}^{2}}F(\phi ),  \label{100t}
\end{equation}%
where $\kappa $ is a constant. This permits to rewrite the action for a $4$%
-dimensional brane non-minimally coupled to a scalar field, immersed in a $5$%
-dimensional spacetime as%
\begin{equation}
S[g,\phi ]=\int d^{4}x\sqrt{-g}\left[ R+\varepsilon RV(\phi )-2\kappa V(\phi
)\right] .  \label{33t}
\end{equation}%
Note that this action was obtained from the Lagrangian (\ref{26}) that does
not consider the presence of a cosmological constant.

The corresponding field equations describing the behavior of the $4$%
-dimensional brane in the presence of the scalar field $\phi $ are given by 
\cite{bra1}, \cite{bra2}

\begin{equation}
G_{\mu \nu }=-\kappa \left( \frac{V}{1+\varepsilon V}\right) g_{\mu \nu },
\label{z1}
\end{equation}%
\begin{equation}
\frac{\partial V}{\partial \phi }\left( 1-\frac{\varepsilon R}{2\kappa }%
\right) =0.  \label{z2}
\end{equation}

In this article we study the relationship between the so-called
Einstein-Chern-Simons gravity compactified to $4$ dimensions and the
scalar-tensor gravity theory. \ It is found that the $4$-dimensional theory
corresponds to a scalar-tensor theory without kinetic term is equivalent to
a $f(R,\phi )$-type theory \cite{1}\textbf{. } In Section $II$ we discuss
about the relationship between the Einstein-Chern-Simons gravity
compactified to $4$ dimensions and the scalar-tensor gravity theory. \ In
Section $III$ we find that field equation (\ref{z1}) for a spherically
symmetric metric, lead to two equations (namely (\ref{11}, \ref{13}))
dependent on the potential $V(\phi )$ present in the action (\ref{33t}),
while the equation (\ref{z2}) gives rise to an equation (namely (\ref{z4}))
which leads to two cases of interest: one case where $\partial V/\partial
r\neq 0$ i.e., $R=2\kappa /\varepsilon $ and a second case where $V$ is a
constant meaning that $1-\varepsilon R/2\kappa \neq 0$. The first case leads
to a non-causal solution and the second case to a spacetime with a
"cosmological constant" inversely proportional to the square of the
compactification radius and to a solution dependent on an arbitrary constant 
$A$. In Section $IV$ we discuss the case when this constant is positive, in
particular, when such constant is equal to the Schwarzschild radius. In this
case the solution is of the type of a Schwarzschild-de Sitter (SdS) black
hole solution.

A procedure used in the context of general relativity \cite{bonda} is
generalized in Section $V$ in order to apply it to a compactified
Chern-Simons gravity. This method is based on the symmetries of parity $%
r\longrightarrow -\tilde{r}$, time reversal $t\longrightarrow -\tilde{t}$
and mass reversal $M\longrightarrow -\tilde{M}$ called $PTM$ symmetry. This
symmetry: $\left( i\right) $ leaves the metric invariant in form, $\left(
ii\right) $ relates the different regions of the extended solution of the
Einstein field equations written in terms of the Kruskal-Szekeres
coordinates \cite{kr}, \cite{sz} and $\left( iii\right) $ reverse the signs
of the Kruskal-Szekeres coordinates. In this Section we will study the
behavior of solutions of the field equations (\ref{z1} and \ref{z2}) under
the aforementioned transformations $PTM$. Finally Concluding Remarks are
presented in Section $VI$. An Appendix is also included, where are
considered details omitted in the main text. We will use $8\pi G=c=1$ units.

\section{Einstein-Chern-Simons gravity and scalar-tensor theory}

It is not difficult to prove that the action (\ref{33t}) is a action for a
scalar-tensor theory for which the kinetic term for the scalar field $\phi $%
\ is zero. In this action the potential for the scalar field determines the
functional form of the nonminimal coupling with Ricci scalar. \ In fact, the
action (\ref{33t}) can be written in the form

\begin{equation}
S\left[ g,\phi \right] =\int d^{4}x\sqrt{-g}\left[ R+f\left( R,\phi \right) %
\right] ,  \label{4'}
\end{equation}%
where $f(R,\phi )=\varepsilon \left[ R-2\kappa /\varepsilon \right] V\left(
\phi \right) ,$ whose field equation is given by (\ref{z1}), which can be
written in the form

\begin{equation}
G_{\mu \nu }=\kappa T_{\mu \nu }=-\kappa \left( \frac{V}{1+\varepsilon V}%
\right) g_{\mu \nu },  \label{5'}
\end{equation}%
which allow us to identify the source, $T_{\mu \nu }$, that generates black
holes solutions which are discussed later. Note that when the $V$ potential
\ is a positive constant, the term $V/(1+\varepsilon V)$\ can be interpreted
as the cosmological constant. In this case the action given in (\ref{4'})
becomes an action for a $f(R)$ theory .

\section{Field equations for a spherically symmetric me\-tric}

In $4$-dimensions a static and spherically symmetric metric can be written
in the form 
\begin{equation}
ds^{2}=-e^{\alpha (r)}dt^{2}+e^{\beta (r)}dr^{2}+r^{2}\left( d\theta
^{2}+\sin ^{2}\theta d\varphi ^{2}\right) ,  \label{8'}
\end{equation}%
where $r,\theta $ and $\varphi $ are the usual spherical polar coordinates.
Introducing (\ref{8'}) in (\ref{z1}), we find the following field equations
(see Appendix)

\begin{eqnarray}
\left[ \frac{-1}{r^{2}}+e^{\alpha }\left( \frac{\alpha ^{\prime }}{r}+\frac{1%
}{r^{2}}\right) \right] \left( 1+\varepsilon V\right) &=&-\kappa V,
\label{11} \\
-\frac{R}{2}\left( 1+\varepsilon V\right) -\left[ \frac{-1}{r^{2}}+e^{\alpha
}\left( \frac{\alpha ^{\prime }}{r}+\frac{1}{r^{2}}\right) \right] \left(
1+\varepsilon V\right) &=&-\kappa V.  \label{13}
\end{eqnarray}

Adding these last equations we obtain%
\begin{equation}
\frac{\varepsilon R}{2\kappa }=\frac{2\varepsilon V}{1+\varepsilon V}.
\label{13'}
\end{equation}

So far we have only used the equations (\ref{z1}), but we have not yet used
the equation (\ref{z2}), which can be written in the form

\begin{equation}
\frac{\partial V}{\partial r}\left( 1-\frac{\varepsilon R}{2\kappa }\right)
=0.  \label{z4}
\end{equation}%
This equation is satisfied when:

\begin{enumerate}
\item[(i)] $\partial V/\partial r\neq 0$, which lead $R=2\kappa /\varepsilon 
$ and therefore $f(R,\phi )=0.$ This means that equation (\ref{4'}) takes
the form of the Einstein-Hilbert action\textbf{\ }%
\begin{equation}
S\left[ g\right] =\int d^{4}x\sqrt{-g}R.  \label{z5}
\end{equation}

\item[(ii)] $\partial V/\partial r=0$, i.e., $V=$ constant $\neq
1/\varepsilon $, which implies $R\neq 2\kappa /\varepsilon $, as can also be
seen from the equation (\ref{13'}), except when $\varepsilon V=1$. This
allows us to state that 
\begin{equation}
f(R,\phi )=\varepsilon \left( R-\frac{2\kappa }{\varepsilon }\right) V(\phi
)\neq 0.  \label{z6}
\end{equation}
\end{enumerate}

Note that it is straightforward to see that when $\varepsilon V=1$\ the
action (\ref{32t}) takes the form 
\begin{equation}
S[g]=2\int d^{4}x\sqrt{-g}\left[ R-\frac{3}{4r_{c}^{2}}\right] ,  \label{z7}
\end{equation}

that is, it takes the form of the Einstein Hilbert action with cosmological
constant. 

\subsection{\textbf{Case where the potential }$V(r)$ is linear in $r$}

Solving the equation (\ref{11}) for the case where $V\left( r\right) =ar+b$,
we find

\begin{eqnarray}
e^{\alpha \left( r\right) } &=&\left[ 1+\frac{\kappa }{\varepsilon }\left( 
\frac{1+\varepsilon b}{\eta }\right) ^{2}\right] \frac{\ln \left(
1+\varepsilon b+\eta r\right) }{\eta r}+\frac{C_{1}}{r}  \notag \\
&&-\frac{\kappa }{3\varepsilon }\left[ r^{2}+\frac{3}{2\eta ^{2}}\left[
2\left( 1+\varepsilon b\right) -\eta r\right] \right] ,  \label{19}
\end{eqnarray}%
where $\eta =a\varepsilon $. \ From here, it is not difficult to see that $%
e^{\alpha \left( r\right) }<0$ which implies that the equation (\ref{19})
leads to a non-causal solution and therefore does not represent a black
hole. This case does not correspond to an $f(R,\phi )$ gravity.

\subsection{ \textbf{Case where the potential }$V(r)$ is constant}

In this case $V^{\prime }=0$ hence (\ref{11}-\ref{13}) takes the form 
\begin{eqnarray}
\left[ \frac{-1}{r^{2}}+e^{\alpha }\left( \frac{\alpha ^{\prime }}{r}+\frac{1%
}{r^{2}}\right) \right] \left( 1+\varepsilon C\right) &=&-\kappa C,
\label{20a} \\
-\left[ \frac{-1}{r^{2}}+\frac{R}{2}+e^{\alpha }\left( \frac{\alpha ^{\prime
}}{r}+\frac{1}{r^{2}}\right) \right] \left( 1+\varepsilon C\right)
&=&-\kappa C,  \label{20b}
\end{eqnarray}%
where $C$ es una constante positiva ($C>0$). \ Adding up (\ref{20a}) and (%
\ref{20b}) we find%
\begin{equation}
R=\frac{4\kappa C}{\varepsilon C+1}.  \label{24}
\end{equation}%
It is not hard to find that the equations (\ref{20a}) and (\ref{20b}) have
the same solution given by%
\begin{equation}
e^{\alpha (r)}=\Phi (r)=1-\frac{A}{r}-Br^{2},  \label{25}
\end{equation}%
where%
\begin{equation}
A=\frac{\tilde{C}}{\left( 1+\varepsilon C\right) }\text{, \ \ \ }B=\frac{%
\kappa C}{3\left( 1+\varepsilon C\right) },  \label{26'}
\end{equation}%
with $C$ and $\tilde{C}$ constants. So the line element takes then the form%
\begin{equation}
ds^{2}=-\left( 1-\frac{A}{r}-Br^{2}\right) dt^{2}+\left( 1-\frac{A}{r}%
-Br^{2}\right) ^{-1}dr^{2}+r^{2}d\Omega ^{2}.  \label{27}
\end{equation}

This results could be also found from the reference \cite{fara1} where were
generalized the Hawking's results of reference \cite{hawk} and established
that "the isolated black holes which are the end-state of collapse are
solutions of scalar-tensor theories if and only if they are solutions of
general relativity".

The line element (\ref{27}) is a $sgn(A)$-dependent solution. So that the $%
\Phi (r)$ function is subject to the discution of $A$, that is, its sign. If 
$A=0$ we obtain the usual solution that leads to a black hole whose event
horizon is a cosmological one. For $A>0$\textbf{\ }we obtain a SdS black
hole if we $A=r_{s}$, $B=\Lambda /3,$ where $r_{s}$ is the Schwarzschild
radius and $\Lambda $ the cosmological constant. This fact is compatible
with the result $R=4\Lambda $ as we can see from (\ref{24}) and (\ref{26'}).
This result matches equation $\left( 28\right) $ from reference \cite{cruz}
in four dimensions with $\Lambda =3B=3/l^{2}.$

The case $A<0$, discussed in Section $IV$, \ leads to a white hole solution
which is obtained applying to the solution with $A>0$, the discrete
coordinate transformation $PT$ accompanied by the transformation $%
A\longrightarrow -\tilde{A}$, which can be interpreted as the mass reversal.

\section{$\mathbf{A>0}$ case: \textbf{SdS black hole solution}}

From (\ref{27}), we write%
\begin{equation}
ds^{2}=-\Phi (r)dt^{2}+\frac{dr^{2}}{\Phi (r)}+r^{2}d\Omega ^{2},  \label{38}
\end{equation}%
where $\Phi (r)$ is given by (\ref{25}). If $A=0$, then the function $\Phi
(r)$ takes the form%
\begin{equation}
\Phi (r)=1-\frac{\Lambda }{3}r^{2},  \label{39}
\end{equation}%
and we have a divergence at $r_{0}=\sqrt{3/\Lambda }$, case that can be seen
as the limit at large $r$ ($dS$ space limit) of the SdS black hole.

Let us consider now the conditions under which the field equations admit
black hole-type solutions. The solution (\ref{38}) with $\Phi (r)$ given by (%
\ref{25}) shows a singular behaviour at $\Phi (r_{0})=0$, i.e.,

\begin{equation}
r_{0}^{3}-\frac{1}{B}r_{0}+\frac{A}{B}=0,  \label{39'}
\end{equation}%
where, $A$ and $B$ are given by (\ref{26'}). \ From (\ref{39'}) we see that
the discriminant is%
\begin{equation}
\Delta =\frac{1}{B^{2}}\left( \frac{4}{B}-27A^{2}\right) =\frac{27\left[
4(1+C\varepsilon )-9\kappa C\tilde{C}^{2}\right] }{C^{3}\kappa ^{3}},
\label{40'}
\end{equation}%
and we must have $\Delta >0$\ if we require real solutions. So

\begin{equation}
A=r_{s}<\frac{2}{3}\sqrt{\frac{1+\varepsilon C}{\kappa C}}.
\end{equation}

Hence (\ref{38}) takes the form\ \ \ 

\begin{equation}
ds^{2}=-\left( 1-\frac{r_{s}}{r}-\frac{\Lambda }{3}r^{2}\right)
dt^{2}+\left( 1-\frac{r_{s}}{r}-\frac{\Lambda }{3}r^{2}\right)
^{-1}dr^{2}+r^{2}d\Omega ^{2},  \label{18'}
\end{equation}%
\ 

where, recalling that $\varepsilon =4\kappa r_{c}^{2}/3$, $\Lambda $ is
given by

\begin{equation}
\Lambda =\frac{R}{4}=\frac{\kappa C}{1+\varepsilon C}.  \label{19'}
\end{equation}%
Choosing $C$ so that $\varepsilon C>>1$ we have

\begin{equation}
\Lambda \approx \frac{3}{4r_{c}^{2}},
\end{equation}%
and if we take into account that $\Lambda \approx 10^{-52}\left[ m^{-2}%
\right] $, we find $r_{c}\approx 0.9a_{0},$ where $a_{0}\approx 10^{26}\left[
m\right] $ is the value of the cosmic scale factor today. In Ref. \cite%
{randall1}, we can read: To conclude, we have found that we can consistently
exist with an infinite fifth dimension, without violating known tests of
gravity. The community is convinced that if an extra dimension played a role
in the universe we live in, it would be difficult to detect given its
"smallness" (theoretical and/or experimental arguments seem to support
this). On the other hand, claiming that said extra dimension is "infinite"
in size seems to be a rather "untenable" claim. We note that under $%
\varepsilon C>>1$ we have $r_{s}<\left( 4/3\sqrt{3}\right) r_{c}$.

We finish this Section highlighting the relation between $\Lambda $ and $%
r_{c}$. We must remember that, according to the action \textbf{(}\ref{33t}%
\textbf{)}, we do not have a cosmological constant. However, $R/4$ plays
such a role.

\section{$A<0$ case: white hole solution}

This case leads to a solution that can be understood as a white hole. To
prove this we will generalize to the case of a Chern-Simons theory, a
procedure introduced in the context of GR \cite{bonda} to construct a white
hole solution from a black hole solution.

From (\ref{39'}) we see that exist a singular behaviour at $\Phi (r_{0})=0$,
i.e., at $r_{0}^{3}=\frac{1}{B}r_{0}-\frac{A}{B},$ such that 
\begin{equation}
27A^{2}>4/B,  \label{35'}
\end{equation}%
i.e., $\Delta <0$. This means that the equation $\Phi (r_{0})=0$ has only a
real and positive root and its discriminant\ tells us that 
\begin{equation}
A^{2}>\frac{4}{9}\frac{1+\varepsilon C}{\kappa C},
\end{equation}%
and the solution for $r_{0}$ is given by

\begin{eqnarray}
r_{0} &=&-\frac{2^{1/3}(1+C\varepsilon )}{\left( -3C^{2}\kappa ^{2}\tilde{C}%
\pm 3\kappa C\tilde{C}\sqrt{C^{2}\kappa ^{2}-\left( 64/9\right) \kappa
C(1+C\varepsilon )^{3}}\right) ^{1/3}}  \notag \\
&\ &-\frac{\left( -3C^{2}\kappa ^{2}\tilde{C}+3\kappa C\tilde{C}\sqrt{%
C^{2}\kappa ^{2}-\left( 64/9\right) \kappa C(1+C\varepsilon )^{3}}\right)
^{1/3}}{2^{1/3}C\kappa }.  \label{69''}
\end{eqnarray}

The solution (\ref{38}) appears to have a singularity at $r=0$ and $r=r_{0}.$
The singularity at $r=r_{0}$ divides the coordinates in two disconnected
patches. The exterior solution with $r>r_{0}$ is the one that is related to
the gravitational fields of massive objects. The interior solution with $%
0\leq r\leq r_{0}$, which contains the singularity at $r=0$, is completely
separated from the outer patch by the singularity at $r=r_{0}$.

The singularity at $r=r_{0}$ is a coordinate singularity. This singularity
arises from a bad choice of coordinates or coordinate conditions. A easiest
way to see that at $r=r_{0}$ is to consider the metric in the
Eddington-Finkelstein (EF) coordinates. For the ingoing EF coordinates $%
\left( v,r,t\right) $, which describe the line element in the $I$-$II$\
regions\textbf{\ (}the form of this regions as well as the regions $III$%
\textbf{\ }and\textbf{\ }$IV$ are similar to that of the case of GR which
are shown in Ref.\cite{chan}), we have%
\begin{equation}
ds^{2}=-\Phi (r)dt^{2}+2dvdr+r^{2}d\Omega ^{2},  \label{68'}
\end{equation}%
which is regular at future horizon and the past horizon is at $v=t+r^{\ast
}=\infty $. \ 

Correspondingly, the outgoing EF coordinates $\left( u,r,t\right) $ which
describe the metric in the $III$-$IV$\ regions has the following form

\begin{equation}
ds^{2}=-\Phi (r)dt^{2}-2dudr+r^{2}d\Omega ^{2},  \label{69'}
\end{equation}%
which is regular at past horizon and the future horizon is at $u\equiv
t-r^{\ast }$, which can be interpreted as a new time coordinate where $%
r^{\ast }$ is given by%
\begin{eqnarray}
r^{\ast } &=&\beta \left[ \ln \left( \frac{r-r_{0}}{\left( r+r_{0}\right) %
\left[ \frac{3(1+C\varepsilon )}{\left( r-r_{0}\right) ^{2}}-\kappa C(1-%
\frac{rr_{0}}{\left( r-r_{0}\right) ^{2}})\right] ^{1/2}}\right) \right. 
\notag \\
&\ &\left. +\frac{\sqrt{3}(2(1+C\varepsilon )-\kappa Cr_{0}^{2})}{r_{0}^{2}%
\sqrt{\kappa C}\sqrt{4\frac{(1+C\varepsilon )}{r_{0}^{2}}-\kappa C}}\text{%
arctanh}\left( \frac{\sqrt{\kappa C}(2r+r_{0})}{\sqrt{3}r_{0}\sqrt{4\frac{%
(1+C\varepsilon )}{r_{0}^{2}}-\kappa C}}\right) \right] ,  \notag \\
&&  \label{70}
\end{eqnarray}%
with 
\begin{equation}
\beta =\frac{r_{0}(1+C\varepsilon )}{(1+C\varepsilon )-C\kappa r_{0}^{2}}.
\label{71}
\end{equation}

The coordinates used in (\ref{68'}) and (\ref{69'}) have the advantage that
they describe the neighbourhood of the surface $r=r_{0}$ in a satisfactory
way. However, the metrics (\ref{68'}) and (\ref{69'}) has still a deficiency%
\textbf{,} analogue to that appears in the Schwarzshild solution of GR. This
deficiency is avoided in the Kruskal coordinates, which describe a
geodesically complete space. Defining 
\begin{equation}
U=-\exp \left( -\frac{u}{2\beta }\right) \text{ \ \ },\text{ \ }V=\exp
\left( \frac{v}{2\beta }\right) ,\text{ \ }  \label{73'}
\end{equation}%
we have that the Kruskal coordinates are given by

\begin{equation}
T=\frac{1}{2}(U+V)\text{ \ \ },\text{ \ }X=\frac{1}{2}(U-V),  \label{83}
\end{equation}%
with which the line element takes the form 
\begin{equation}
ds^{2}=-\tilde{\Phi}(r)\left( dT^{2}-dX^{2}\right) +r^{2}d\Omega ^{2},
\label{85}
\end{equation}%
where%
\begin{equation}
\tilde{\Phi}(r)=\frac{4r_{0}(1+\varepsilon C)}{(1+\varepsilon C)-\kappa
Cr_{0}^{2}}\Phi (r)\exp \left( -\frac{r}{r_{0}}\frac{(1+\varepsilon
C)-\kappa Cr_{0}^{2}}{(1+\varepsilon C)}\right) ,\quad r>0.  \label{79}
\end{equation}

Following the usual procedure we construct a spacetime diagram in the $%
\left( T,X\right) $ plane (with $\theta ,\varphi $ suppressed) known as the
Kruskal diagram, which represents the extended spacetime corresponding to
the metric (\ref{85}), whose form is similar to that of the case of GR (see
Ref. \cite{chan} where the aforementioned regions $I,II,III,IV$\ are
shown).\ 

Following the reference \cite{bonda}, the next step is to continue the
metric in the region of the additive inverse of $A$ carrying out the
substitution $A\longrightarrow -\tilde{A}$, with$\tilde{A}>0$, which must be
complemented with the transformation $t\longrightarrow -\tilde{t}$, $%
r\longrightarrow -\tilde{r}$, $r_{c}\longrightarrow -\tilde{r}_{c}$, because
it is necessary to preserve the dynamics. From (\ref{26'}) we can see that
if $A\longrightarrow -\tilde{A}$, then $\tilde{C}\longrightarrow -\bar{C}$
with $\bar{C}$ $>0$, so that under this transformation, (\ref{69''})
transforms as

\begin{eqnarray}
r_{0} &\longrightarrow &-\tilde{r}_{0}=-\frac{2^{1/3}(1+C\varepsilon )}{%
\left[ -\left( -3C^{2}\kappa ^{2}\bar{C}\pm 3\kappa C\bar{C}\sqrt{%
C^{2}\kappa ^{2}-\left( 64/9\right) \kappa C(1+C\varepsilon )^{3}}\right) %
\right] ^{1/3}}  \notag \\
&\ &-\frac{\left[ -\left( -3C^{2}\kappa ^{2}\bar{C}+3\kappa C\bar{C}\sqrt{%
C^{2}\kappa ^{2}-\left( 64/9\right) \kappa C(1+C\varepsilon )^{3}}\right) %
\right] ^{1/3}}{2^{1/3}C\kappa }.  \label{86}
\end{eqnarray}

This means that $A\longrightarrow -\tilde{A}$ (i.e., $\tilde{C}%
\longrightarrow -\bar{C}$) implies $r_{0}\longrightarrow -\tilde{r}_{0}$. On
the other hand under this transformation it is straightforward to see that $%
\tilde{\beta}=-\beta $ and that the Edington Finkelstein coordinate $%
v=t+r^{\ast }$ takes the form

\begin{eqnarray}
\tilde{v} &=&-\tilde{t}-\left\{ \tilde{\beta}\ln \left( \frac{\left( \tilde{r%
}-\tilde{r}_{0}\right) }{\left( \tilde{r}+\tilde{r}_{0}\right) \left[ \frac{%
3(1+C\varepsilon )}{\left( \tilde{r}-\tilde{r}_{0}\right) ^{2}}-\kappa C(1-%
\frac{\tilde{r}\tilde{r}_{0}}{\left( \tilde{r}-\tilde{r}_{0}\right) ^{2}})%
\right] ^{1/2}}\right) \right.  \notag \\
&\ &\left. +\tilde{\beta}\frac{\sqrt{3}(2(1+C\varepsilon )-\kappa C\tilde{r}%
_{0}^{2})}{\tilde{r}_{0}^{2}\sqrt{\kappa C}\sqrt{4\frac{(1+C\varepsilon )}{%
\tilde{r}_{0}^{2}}-\kappa C}}\text{arctanh}\left( \frac{\sqrt{\kappa C}(2%
\tilde{r}+\tilde{r}_{0})}{\sqrt{3}\tilde{r}_{0}\sqrt{4\frac{(1+C\varepsilon )%
}{r_{0}^{2}}-\kappa C}}\right) \right\} -2i\pi \tilde{\beta},  \notag \\
\tilde{v} &=&-\tilde{t}-\tilde{r}^{\ast }-2i\pi \tilde{\beta}=-\nu -2i\pi 
\tilde{\beta},  \label{87}
\end{eqnarray}

where $\tilde{r}^{\ast }$ is given by the expression in parentheses brace.

In the same way it is found that the Eddington-Finkelstein coordinate $%
u=-t+r^{\ast }$ transform as

\begin{equation}
u=-\tilde{t}+\tilde{r}^{\ast }+2i\pi \tilde{\beta}=-u+2i\pi \tilde{\beta}.
\label{88}
\end{equation}

Equations (\ref{87}) and (\ref{88}) allow finding the continuation of
equations (\ref{73'}) in the region of the additive inverse of $A,$ which
can be interpreted as the negative mass region. In fact, under the
transformation $A\longrightarrow -\tilde{A}$ (i.e., $\tilde{C}%
\longrightarrow -\bar{C}$), $t\longrightarrow -\tilde{t}$, $r\longrightarrow
-\tilde{r}$, $r_{0}\longrightarrow -\tilde{r}_{0}$, the equations (\ref{73'}%
) transform as 
\begin{eqnarray}
U &\longrightarrow &\tilde{U}=\exp \left( -\frac{\tilde{u}}{2\tilde{\beta}}%
\right) =-U,  \notag \\
V &\longrightarrow &\tilde{V}=-\exp \left( \frac{\tilde{v}}{2\tilde{\beta}}%
\right) =-V.
\end{eqnarray}%
It is straightforward to see that this transformation implies that the
Kruskal coordinates transform as

\begin{eqnarray}
T &\longrightarrow &\tilde{T}=\frac{1}{2}(\tilde{U}+\tilde{V})=-T,  \notag \\
\tilde{X} &=&\frac{1}{2}(\tilde{V}-\tilde{U})=-X,
\end{eqnarray}%
that is, the inversion of the Kruskal coordinates. As in the case of general
relativity, studied in \cite{bonda}, the spacetime with the variables $(%
\tilde{U},\tilde{V})$, found using the transformations $A\longrightarrow -%
\tilde{A}$ (i.e., $\tilde{C}\longrightarrow -\bar{C}$), $t\longrightarrow -%
\tilde{t}$, $r\longrightarrow -\tilde{r}$, $r_{0}\longrightarrow -\tilde{r}%
_{0}$, can be understood as a space corresponding to a spacetime of the a
white hole. \ This means that region $I$ with coordinates $(\tilde{U},\tilde{%
V})$ corresponds to region $III$ when we use coordinates $\left( U,V\right) $%
.

In the same way, region $IV$, which in the usual coordinates corresponds to
a white hole, is transformed into region $II$ which corresponds, in the new
coordinates, to a black hole. Obviously, region $II$, which in the usual
coordinates corresponds to a black hole, will be transformed into region $IV$
which, in the new coordinates, corresponds to a white hole.

The interesting thing about this result is that the solutions for $A<0$ and $%
A>0$ are related through the transformation $A\longrightarrow -\tilde{A}$,
with $\tilde{A}>0$, $t\longrightarrow -\tilde{t}$, $r\longrightarrow -\tilde{%
r}$, $r_{0}\longrightarrow -\tilde{r}_{0}$.

\section{\textbf{Concluding Remarks }}

We have shown that the so-called $5$-dimensional generalized Poincar\'{e}
Chern-Simons gravity (also called Einstein-Chern-Simons gravity)
compactified to $4$-dimensions is equivalent to a scalar-tensor theory
without kinetic term wich correspond to an $f(R,\phi )$-type theory. We find
that field equation (\ref{z1}) for a spherically symmetric metric, lead to
two equations (namely (\ref{11}, \ref{13})) dependent on the potential $%
V(\phi )$ present in the action (\ref{33t}), while the equation (\ref{z2})
gives rise to an equation (namely (\ref{z4})) which leads to two cases of
interest: the first case occurs when the potential $V$ satisfies the
condition $\partial V/\partial r\neq 0$, which lead $R=2\kappa /\varepsilon $
and therefore $f(R,\phi )=0,$ which means that this case does not correspond
to an $f(R,\phi )$ gravity and the solution to the corresponding field
equations is of a non-causal nature and therefore does not represent a black
hole.

\textbf{\ }The second case occurs when the potential $V$ satisfies the
condition $\partial V/\partial r=0$, i.e., $V=$ constant $\neq 1/\varepsilon 
$, which implies $R\neq 2\kappa /\varepsilon $ and therefore $f(R,\phi )\neq
0$ except when $\varepsilon V=1$. The solution to the corresponding field
equations lead to a spacetime with a "cosmological constant" inversely
proportional to the square of the compactification radius and to a solution
dependent on an arbitrary constant $A$. When this constant is positive, in
particular, when such constant is equal to the Schwarzschild radius we
obtain a solution of the type of a Schwarzschild-de Sitter (SdS) black hole
solution.

A procedure used in the context of general relativity \cite{bonda} was
generalized and applied to a compactified Chern-Simons gravity. This method
is based on the symmetries of parity $r\longrightarrow -\tilde{r}$, time
reversal $t\longrightarrow -\tilde{t}$ which together correspond to the
well-known symmetry\textbf{\ }$PT$ accompanied \ by the mass reversal
transformation $M\longrightarrow -\tilde{M}$. The symmetry $PT$ plus the
mass reversal transformation is known as $PTM$ symmetry. This symmetry
leaves the metric invariant in form and allow us relates the different
regions of the solution of the field equations written in terms of the
Kruskal-Szekeres coordinates \cite{kr}, \cite{sz}. \ We have shown that the
application of the $PTM$ transformation to the solution for $A>0$ leads to a
solution for $A<0$ equivalent to an SdS white hole solution. The change of
sign in the mass can be understood as a consequence of the requirement of
the invariance of the metric under the $PTM$ symmetry.

\section{Appendix: \textbf{Field equations for a spherically symmetric
me\-tric}}

We consider the field equations (\ref{z1}) and (\ref{z2})

\begin{equation}
G_{\mu \nu }\left( 1+\varepsilon V\right) =-\kappa g_{\mu \nu }V,
\label{z1'}
\end{equation}%
\begin{equation}
\frac{\partial V}{\partial \phi }\left( 1-\frac{\varepsilon R}{2\kappa }%
\right) =0.  \label{z2'}
\end{equation}

In $4$-dimensions a static and spherically symmetric metric can be written
in the form 
\begin{equation}
ds^{2}=-e^{\alpha (r)}dt^{2}+e^{\beta (r)}dr^{2}+r^{2}\left( d\theta
^{2}+\sin ^{2}\theta d\varphi ^{2}\right) ,  \label{A1}
\end{equation}%
so that the non-zero Christoffel symbols, the Ricci tensor, the scalar
curvature and the Einstein tensor are given by 
\begin{eqnarray}
\Gamma _{00}^{1} &=&\frac{\alpha ^{\prime }}{2}e^{\alpha -\beta }\text{ \ },%
\text{ }\Gamma _{11}^{1}=\frac{\beta ^{\prime }}{2}\text{ \ },\text{ }\Gamma
_{22}^{1}=-re^{-\beta },  \notag \\
\Gamma _{12}^{2} &=&r^{-1}=\Gamma _{21}^{2}\text{ \ },\text{ }\Gamma
_{33}^{2}=-\sin \theta \cos \theta \text{ \ },\text{ }\Gamma _{23}^{3}=\cot
\theta ,\text{ }  \notag \\
\Gamma _{01}^{0} &=&\Gamma _{10}^{0}=\frac{\alpha ^{\prime }}{2}\text{ \ },%
\text{ }\Gamma _{13}^{3}=r^{-1}=\Gamma _{31}^{3}\text{ \ },\text{ }\Gamma
_{33}^{1}=-re^{-\beta }\sin ^{2}\theta ,  \label{A2}
\end{eqnarray}

\begin{eqnarray}
R_{00} &=&\frac{e^{\alpha -\beta }}{2}\left( \alpha ^{\prime \prime }+\frac{%
\alpha ^{\prime 2}}{2}-\frac{\alpha ^{\prime }\beta ^{\prime }}{2}+\frac{%
2\alpha ^{\prime }}{r}\right) ,  \notag \\
R_{11} &=&-\frac{1}{2}\left( \alpha ^{\prime \prime }-\frac{\alpha ^{\prime
}\beta ^{\prime }}{2}-\frac{2\beta ^{\prime }}{r}+\frac{\alpha ^{\prime 2}}{2%
}\right) ,  \notag \\
R_{22} &=&1-e^{-\beta }\left( 1+\frac{r}{2}\left( \alpha ^{\prime }-\beta
^{\prime }\right) \right) ,  \notag \\
R_{33} &=&R_{22}\sin ^{2}\theta ,  \notag \\
R &=&\frac{2}{r^{2}}-e^{-\beta }\left( \alpha ^{\prime \prime }+\frac{\alpha
^{\prime 2}}{2}-\frac{\alpha ^{\prime }\beta ^{\prime }}{2}+\frac{2}{r}%
\left( \alpha ^{\prime }-\beta ^{\prime }\right) +\frac{2}{r^{2}}\right) ,
\label{A3}
\end{eqnarray}
\begin{eqnarray}
G_{00} &=&\frac{e^{\alpha }}{r^{2}}+e^{\alpha -\beta }\left( \frac{\beta
^{\prime }}{r}-\frac{1}{r^{2}}\right) ,  \notag \\
G_{11} &=&\frac{\alpha ^{\prime }}{r}+\frac{1}{r^{2}}-\frac{e^{\beta }}{r^{2}%
},  \notag \\
G_{22} &=&-\frac{e^{-\beta }}{2}\left( \frac{\alpha ^{\prime }\beta ^{\prime
}}{2}r^{2}-\frac{\alpha ^{\prime 2}}{2}r^{2}-\alpha ^{\prime \prime
}r^{2}-r\left( \alpha ^{\prime }-\beta ^{\prime }\right) \right) ,  \notag \\
G_{33} &=&G_{22}\sin ^{2}\theta ,  \label{A4}
\end{eqnarray}%
where $r,\theta $ and $\varphi $ are the usual spherical polar coordinates
and the "prime"\ denotes derivative with respect to $r$. Introducing (\ref%
{A1},\ref{A4}) in (\ref{z1}) we find

\begin{eqnarray}
\left[ \frac{e^{\alpha }}{r^{2}}+e^{\alpha -\beta }\left( \frac{\beta
^{\prime }}{r}-\frac{1}{r^{2}}\right) \right] \left( 1+\varepsilon V\right)
&=&\kappa e^{\alpha }V,  \label{uno} \\
\left[ \frac{\alpha ^{\prime }}{r}+\frac{1}{r^{2}}-\frac{e^{\beta }}{r^{2}}%
\right] \left( 1+\varepsilon V\right) &=&-\kappa e^{\beta }V,  \label{dos} \\
\left[ 1-\frac{r^{2}}{2}R-\frac{e^{-\beta }}{2}\left[ \left( \alpha ^{\prime
}-\beta ^{\prime }\right) r+2\right] \right] \left( 1+\varepsilon V\right)
&=&-\kappa r^{2}V,  \label{tres}
\end{eqnarray}

\begin{equation}
\left[ 1-\frac{r^{2}}{2}R-\frac{e^{-\beta }}{2}\left[ \left( \alpha ^{\prime
}-\beta ^{\prime }\right) r+2\right] \right] \left( 1+\varepsilon V\right)
\sin ^{2}\theta =-\kappa r^{2}V\sin ^{2}\theta .  \label{cuatro}
\end{equation}

We have seen that when $l\longrightarrow 0$, i. e., $V\rightarrow 0$, the
equations (\ref{z1}),(\ref{z2}) lead to Einstein's equations. This means
that the equations (\ref{uno}-\ref{cuatro}) must lead, in this limit, to the
corresponding Einstein field equations for a spherically symmetric metric.
For this limit to be fulfilled, it must be satisfied that $\alpha (r)=-\beta
(r)$. \ \ Multiplying (\ref{uno}) by $-e^{-\alpha }$, (\ref{dos}) by $%
e^{-\beta }$, and (\ref{tres}) by $1/r^{2}$, and using $\alpha =-\beta $, we
find the equations (\ref{11}, \ref{13}).

\textbf{Acknowledgements}

This work was supported in part by\textit{\ }FONDECYT Grants\textit{\ }No.%
\textit{\ }1180681 and No.\textit{\ }1211219 from the Government of Chile.
One of the authors (CR) was supported by ANID fellowship No. 2219610 from
the Government of Chile and from Universidad de Concepci\'{o}n, Chile. \ One
of the authors (PS) thanks Jos\'{e} D\'{\i}az for interesting comments about
the field equations.


\begin{thebibliography}{99}
\bibitem{salg1} F. Izaurieta, P. Minning, A. P\'{e}rez, E. Rodriguez, P.
Salgado,\textit{\ Phys. Lett. B} 678 (2009) 213.

\bibitem{salg2} F. Izaurieta, E. Rodriguez, P. Salgado, \textit{Jour. Math.
Phys.} 47 (2006) 123512.

\bibitem{salg3} F. Izaurieta, A. P\'{e}rez, E. Rodriguez, P. Salgado, 
\textit{Jour. Math. Phys.} 50 (2009) 073511.

\bibitem{salga1} C. A. Quinzacara and P. Salgado,\textit{\ Phys. Rev. D} 85
(2012) 124026.

\bibitem{bra1} R. D\'{\i}az, F. G\'{o}mez, M. Pinilla and P. Salgado, Eur.
Phys. J. C80 (2020) 546.

\bibitem{bra2} F. G\'{o}mez, S. Lepe and P. Salgado, Eur. Phys. J. C 81
(2021) 1.

\bibitem{1} The equivalence between a scalar-tensor theory without kinetic
term and the $f(R)$\ theory was\ studied in Refs. \cite{fara1}, \cite{cruz}

\bibitem{randall} L. Randall and R. Sundrum, Phys. Rev. Lett. 83 (17),
(1999) 3370.

\bibitem{randall1} L. Randall and R. Sundrum, Phys. Rev. Lett. 83 (17),
(1999) 4690.

\bibitem{weinberg} S. Weinberg, Gravitation and Cosmology, John Wiley \&
Sons, 1972.

\bibitem{fara1} T.P. Sotiriou and V. Faraoni, Phys. Rev. Lett. 108 (2012)
081103.

\bibitem{cruz} A. de la Cruz-Dombriz, A. Dobado, and A.L. Maroto, Phys. Rev.
D80 (2009) 124011.

\bibitem{teys} P. Teyssandier and Ph. Tourrenc, J. Math. Phys. 24 (1983) 2793

\bibitem{bonda} S. Bondarenko, "Negative mass scenario and Schwarzschild
spacetime in general relativity", Mod. Phys. Lett. A 34 (2019) 1950084

\bibitem{kr} M. D. Kruskal, "Maximal extension of Schwarzschild metric".
Phys. Rev. 119 (1960) 1743

\bibitem{sz} G. Szekeres, "On the singularities of a Riemannian manifold".
Publ. Math. Debrecen 7 (1960) 285

\bibitem{chan} S. Chandrasekhar, "The Mathematical Theory of Black Holes".
Clarendon Press. Oxford, UK, 1983

\bibitem{fro} V.P. Frolov, I.D. Novikov, "Black Holes Physics". Kluwer
Academic Publishers. Dordrecht, The Netherlands, 1998.

\bibitem{hawk} S.W. Hawking, Commun. Math. Phys. 25 (1972) 152
\end{thebibliography}
\end{document}